\documentclass[usenatbib]{mn2e}
\usepackage{color} 
\usepackage{graphicx}

\usepackage{graphics}
\usepackage{epsfig}
\usepackage{natbib}

\begin{document}

\title[ IllustrisTNG and Illustris simulation comparison of the CGM] 
{The morphology and kinematics of the gaseous circumgalactic medium of Milky Way mass galaxies
-- II. comparison of IllustrisTNG and Illustris simulation results}

\author [G.Kauffmann et al.] {Guinevere Kauffmann$^1$\thanks{E-mail: gamk@mpa-garching.mpg.de},
Dylan Nelson$^1$, Sanchayeeta Borthakur$^2$,
Timothy Heckman$^3$,
\newauthor Lars Hernquist$^4$, Federico Marinacci$^5$, R\"udiger Pakmor$^1$, Annelisa Pillepich$^6$\\ 
$^1$Max-Planck Institut f\"{u}r Astrophysik, 85741 Garching, Germany\\
$^2$School of Earth and Space Exploration, Arizona State University, Tempe, AZ 85287, USA\\
$^3$Center for Astrophysical Sciences, Department of Physics and Astronomy, Johns Hopkins University, Baltimore, MD 21218, USA\\
$^4$Harvard-Smithsonian Center for Astrophysics, 60 Garden Street, Cambridge, MA 02138, USA\\
$^5$Kavli Institute for Astrophysics and Space Research, Massachusetts Institute of Technology, Cambridge, MA 02139, USA\\ 
$^6$Max-Planck-Institut für Astronomie, Königstuhl 17, D-69117 Heidelberg, Germany}

\maketitle

\begin{abstract} 
We have carried out a controlled comparison of the structural and kinematic
properties of the circumgalactic medium (CGM) around Milky Way mass
galaxies in the Illustris and IllustrisTNG simulations. Very striking
differences are found. At z=0, gas column density and temperature profiles at large
radii ($\sim 100$ kpc) correlate strongly with disk gas mass fraction in
Illustris, but not in TNG. The neutral gas at large radii is preferentially
aligned in the plane of the disk in TNG, whereas it is much more isotropic
in Illustris.  The vertical coherence scale of the rotationally
supported gas in the CGM is  linked to the gas mass fraction of the galaxy
in Illustris, but not in TNG.  A tracer particle analysis allows us to show
how these differences can be understood as a consequence of the different
sub-grid models of feedback in the two simulations. A  study of
spatially matched galaxies in the two simulations shows that in TNG, 
feedback by supernovae and AGN helps to create an extended smooth reservoir of hot
gas at high redshifts, that then cools to form a thin,
rotationally-supported disk at later times. In Illustris, AGN dump heat in
the form of hot gas bubbles that push diffuse material at large radii out
of the halo. The disk is formed by accretion of colder, recycled material, and this results
in more vertically extended gas distributions above and below the Galactic plane.  We conclude
that  variations in the  structure of gas around Milky Way mass galaxies are a 
sensitive  probe of feedback physics in simulations and are worthy of more observational
consideration in future.
\end {abstract}
\begin{keywords}  galaxies:haloes; galaxies: formation; galaxies: structure ; galaxies: gas content     
\end{keywords}

\section{Introduction}

The thin stellar disk of our  Milky Way is believed to have formed over the last
9 Gyr, i.e.  since a redshift of $\sim 1$, making it the youngest stellar component of
our Galaxy (see Rix \& Bovy 2013 for a recent review). The current ongoing star
formation in the Milky Way would exhaust the available molecular gas supply in
just 1-2 Gyr, which means that disk gas must be replenished from an outer
reservoir. Outer gas reservoirs may exist in the form of an extended disk of
atomic gas at lower densities, or as diffuse gas in a more spherical halo that
cools and accretes onto the main disk over time (Bauermeister, Blitz \& Ma 2010)

Studies of the circum-galactic medium (CGM) around galaxies provide a means of
understanding gas accretion processes in more detail. The CGM is generally
defined as gas that is outside the main disk of a galaxy, but within the virial
radius of its surrounding dark matter halo. In our own Milky Way and in the most
nearby galaxies, it is possible to study individual clouds in the CGM
(see Putman, Peek \& Joung 2012 for a review). In more distant galaxies, 
the gas may either be observed in
absorption, most commonly in the spectra of high redshift quasars whose
sightlines pass within a few hundred kiloparsecs of a galaxy at lower redshift (e.g. Zhu et al 2014),
or in emission in the form of X-ray emitting gas at high temperatures, as
extra-planar, diffuse ionized gas around low-redshift galaxies (e.g. Jones et al
2017), or as giant Ly$\alpha$ emitting halos around quasars at high redshifts
(Martin et al 2016; Arrigoni-Battaia et al 2018).  One major complication in the
interpretation of such observations is that CGM gas is likely to consist of a
mixture of outflowing gas that has been heated by supernovae in the disk, and
inflowing gas that will fuel future star formation.

The interplay between inflowing and outflowing gas on CGM observables can be
studied using simulations of galaxy formation that calculate the hydrodynamical
evolution of gas under the influence of the gravitational field of an evolving
dark matter density field in cosmological volumes of the Universe. These
simulations have now reached sufficient particle number and mass resolution to
track the formation and evolution of hundreds of galaxies of Milky Way mass
within volumes of 100 Mpc$^3$ with spatial resolution of around 1 kpc (e.g.
Vogelsberger et al 2014a; Genel et al 2014;  Dubois et al 2014;
Khandai et al 2015; Nelson et al 2015a; Schaye et al
2015). Recent work has demonstrated that the radial extent
and covering fraction of Ly$\alpha$ and metal absorption line gas is extremely
sensitive to the detailed implementation of feedback from supernovae and AGN in
numerical hydrodynamical models (Suresh et al 2015).  This is particularly true
of tracers of transient warm gas, such as OVI, which are sensitive probes of
recent energy ejection processes (Nelson et al 2018b).

Attempts to use CGM observables to elucidate  gas accretion mechanisms are less
advanced.  Stewart et al (2011) (see also Danovich et al 2014) 
suggested  high angular momentum gas co-rotating
with the galactic disk as a clear and unambiguous signature of cosmological gas
accretion onto galaxies. Their analysis of simulations of Milky Way mass
galaxies  at z=2 suggested that the co-rotating material would form a thickened,
warped extended structure that should be detectable as background object
absorption lines offset from the central galaxy's systemic velocity in a single
direction. Evidence has been found that the population of MgII absorbers located
close to the galactic disk plane have Doppler shifts with the same sign as the
galactic rotation, lending credence to this picture (Kacprzak et al 2012; Ho et al 2017;
Martin et al 2019).  The
kinematics of gaseous halos around quasars can be studied in considerable
detail, because the strong radiation field emitted by the accreting black hole
ionizes the gas out to large distances (Liu et al 2013). Integral field unit spectroscopy of
Ly$\alpha$ emitting halos around quasars at redshift 2-3 have revealed
large-scale coherent rotation-like patterns spanning 300 km/s with a large
velocity dispersion, which are interpreted as a signature of the inspiralling
accretion of substructures within the quasar's host halo (Arrigoni-Battaia et al
2018).

Although it has been argued that high angular momentum CGM gas co-rotating with
the disk is a generic qualitative prediction of the $\Lambda$ Cold Dark Matter
(LCDM) cosmological paradigm (Stewart et al 2017; Stevens et al 2017;
Oppenheimer 2018), it is not yet understood how robustly simulations can predict
the detailed morphology, kinematics and temperature structure of this gas, and
how such properties are expected to vary as a function of galaxy properties.

Kauffmann, Borthakur \& Nelson (2016, hereafter KBN16) carried out a comparison
of the CGM for galaxies with different disk cold gas mass fractions using the
Illustris simulations. This was motivated by the observational findings of
Borthakur et al (2015), showing strong correlation (99.8\% confidence) between the gas fraction
of the galaxy and the impact-parameter-corrected Ly$\alpha$ equivalent width of absorbers 
in the spectra of background quasi-stellar objects at impact parameters of 63-231 kpc.  
KBN16  found that the Illustris neutral hydrogen column
densities in the CGM increased as a function of disk gas mass fraction out to
radii of 100 kpc.  The neutral hydrogen was found to rotate coherently about the
centre of the galaxy with a maximum rotational velocity of around 200 km/s. In
gas-rich galaxies, the average {\em vertical} coherence length of the rotating
gas was 40 kpc, compared to only 10 kpc in gas-poor galaxies.

In this paper, we test the robustness of these results to changes in the
subgrid-models for supernovae and AGN feedback by undertaking  an identical
analysis for the new IllustrisTNG simulation. We find that clear trends in CGM
properties with disk gas mass fraction are almost entirely absent in
IllustrisTNG. In order to understand why the two simulations give such different
results, we carry out an analysis of the heating/cooling and inflow/outflow
patterns of circumgalactic gas as a function of radius in the halo for a set of
Milky Way mass galaxies with different disk gas mass fractions. Because the two
simulations are run using identical initial conditions, we can carry out the
analysis for a spatially matched set of halos, thus eliminating differences that
may arise as a result of different halo formation histories. In addition, we
note that the two simulations are carried out using the same code AREPO
(Springel et al 2010), thus eliminating differences due to different
numerical schemes.

Our paper is organized as follows. In section 2, we summarize the main changes
to the physical prescriptions between the Illustris and IllustrisTNG simulations
relevant for this study. In section 3, we directly compare the radial profiles,
morphology and rotational kinematics of the gas in the two simulations in three
bins of disk gas mass fraction. In section 4, we describe how tracer particles
can be used to track the temperature evolution and motions of gas between
different simulation snapshots. We present a comparison of how heating/cooling
as well as inflow/outflow patterns differ between matched galaxies in the two
simulations.  In section 5, we summarize our main results and attempt to
elucidate how and why changes to the feedback prescriptions lead to very
different CGM structure and morphology around present-day galaxies.

\section {Simulations and Samples}

We analyze two simulations herein.
First, we make use of data from the Illustris-1 simulation
(Vogelsberger et al (2014a,b),Genel et al 2014), the highest resolution of the publicly
released
simulation boxes with a volume of
$10^{6.5}$ Mpc$^3$, dark matter and gas particles masses of $6.3 \times 10^6
M_{\odot}$ and $1.6\times10^6 M_{\odot}$, and dark matter and gas gravitational
softening lengths of 1.4 and 0.7 kpc.
Second, we include simulations from the
IllustrisTNG project (Pillepich et al 2018, Springel et al 2018, Naiman 
et al 2018, Nelson et al 2018a, Marinacci et al 2018), which includes three
distinct simulation volumes: TNG50, TNG100, and
TNG300. Here, we make use of TNG100, for which  the baryon mass resolution is
$1.4 \times 10^6 M_{\odot}$.  
For the complete numerical
details about this run, see Table A2 of Nelson et al. (2018a). An overview of the
differences between the TNG galaxy formation physics model and the original
Illustris simulation model, including the fiducial values of all parameters, is
given in Table 1 of Pillepich et al (2018). The main differences that are likely
to  impact the CGM are differences in the treatment of galactic winds and black
hole feedback, so we briefly summarize them here.

\subsection {Galactic winds} In TNG, wind particles are launched isotropically,
rather than in a direction perpendicular to the disk. Tests show that this
change has little influence on the morphology of the galactic wind (Figure 5 of
Pillepich et al 2018). 
The change that is likely to impact CGM properties most
strongly is the revised scaling of the wind particle speed with galaxy/halo
properties.  In Illustris, the wind particle speed is assumed to scale with the
local, one-dimensional dark matter velocity dispersion -- the prescription
follows the implementation of feedback in Oppenheimer \& Dav\'e (2006). In
IllustrisTNG, the wind speed is assumed to scale not only with $\sigma_{DM}$ but
also with redshift, in such a way that the wind speed is redshift-independent at
fixed halo mass. 
Unlike Illustris, TNG also includes a minimum wind velocity, so that the wind
injection speed never drops below 350 km/s in any halo at any redshift. This
increases the effectiveness of feedback in low mass galaxies and at higher
redshifts. Wind mass loading factors also evolve in opposite
directions at fixed halo mass in TNG and Illustris.  
Finally, in addition to the changes in wind injection velocity, TNG
includes two additional changes affecting how the available wind energy is
distributed: i) some given fraction of this energy is thermal, ii) the wind
energy depends on the metallicity of the starforming gas cell, such that
galactic winds are weaker in higher metallicity environments. In a dark matter
halo of $\sim 10^{12} M_{\odot}$, this means that the wind mass loading factors
at injection are the same in Illustris and IllustrisTNG at $z \sim 2-3$, but a factor of two
lower in TNG at $z=0$.

In summary, for Milky Way type halos, galactic winds have similar properties in
the two simulations at high redshifts, but at low redshifts, TNG winds are
injected with higher velocities, but with lower mass loading factors.

\subsection {Black hole feedback}

Both Illustris and IllustrisTNG include two modes of AGN feedback. The mode
operating at high accretion rates, the so-called quasar mode, is similar in the
two simulations. A fraction of the radiative energy released by the accreted gas
couples thermally to nearby gas within a radius that contains some fixed amount
of mass. For  low-activity states of the black hole, a ``bubble model'',  a form
of mechanical radio-mode AGN feedback following Sijacki et al. (2007) is
implemented in Illustris.  Bubbles of hot gas with radius $\sim$ 50 kpc and total
energy $ \sim 10^{60}$ erg and volume density $ \sim 10^4 M_{\odot}$ kpc$^{-3}$ are placed
into the halo at distances of $\sim 100$ kpc from their centres. The radio-mode
feedback efficiency provided by the bubbles is assumed to be a fixed fraction of
the rest mass energy of the gas accreted by the black hole. The main problem
with this model as implemented in Illustris, was that a significant fraction of the gas in halos of masses
of order $10^{13} M_{\odot}$ was expelled beyond the virial radius of the halo
by the present day, violating constraints on the observed X-ray luminosities of
groups and clusters (Genel et al 2014). In addition, the Illustris bubble model did not
quench ongoing star formation in the central galaxies of massive halos.  These
problems motivated a switch to a kinetic feedback model in IllustrisTNG
(Weinberger et al 2017).

In both Ilustris and IllustrisTNG, the Eddington ratio is used as the criterion
for deciding the accretion state of the black hole. IllustrisTNG includes an
additional dependence on black hole mass that acts to push massive black holes
into the kinetic mode at higher Eddington ratios compared to lower mass black
holes. The ``pivot'' black hole mass where the kinetic mode becomes dominant is
around $10^8 M_{\odot}$. Unlike in the high-accretion state, momentum but no
thermal energy is input into gas cells in the neighbourhood of the black hole.
Note that the  momentum is added in a random direction 
when averaged in time across multiple feedback events  -- 
surprisingly, the 
kinetic mode at low Eddington rates does  produce coherent gas flows, as might be expected in the
case of a large-scale relativistic jet interacting with the CGM (see Nelson et al 2019, in
prep).  As in the
bubble feedback model, the kinetic feedback mode is discretized by imposing a
minimum energy that needs to accumulate  before the
feedback energy is released. The adopted energy threshold is assumed to scale
with the square of the dark matter velocity dispersion multiplied by the gas
mass in the feedback region. This scaling was chosen to ensure that in
IllustrisTNG, the specific energy of the AGN wind within individual
injections  did not significantly exceed
the specific binding energy of the halo, resulting in minimal gas escape from
the halo and changes to the thermodynamic state of the gas on large scales.

\subsection {Magnetic Fields}
We note that another addition in the TNG model is  the inclusion of
magnetic fields (Pakmor, Bauer \& Springel 2011). Galaxies in runs with and
without magnetic fields have been found to differ somewhat in their gas masses
and sizes (Pillepich et al 2018). Within the disk, magnetic fields appear to
have  little dynamical impact on structures such as spiral arms, bulges and bars
(Pakmor \& Springel (2013); Pakmor et al.
(2017)). The impact of the magnetic field on the structure of the CGM has not
yet been examined in detail.

\subsection {Samples}

We now describe the two samples of halos analyzed in this paper. The first 
of these is used to generate Figures 1-6 and pertain to the analysis
of the statistical properties of the CGM as a function of
galaxy gas mass fraction. The second sample is used
for the comparative  analysis of gas flows and heating/cooling and is used to 
generate Figures 7-14.  

\subsubsection {Statistical analysis of CGM properties as a function of galaxy gas
mass fraction}

The analysis of Milky Way type galaxies in KBN16 focused on a sample of galaxies
with stellar masses (calculated within twice the half mass radius) in the range
6-8 $\times 10^{10} M_{\odot}$ residing in subhalos with dark matter masses in
the range 1$-$2 $\times 10^{12} M_{\odot}$. The sample was divided into four bins
in gas mass fraction $f_g=M_{gas}/M_{stars}$, where $M_{gas}$ was again evaluated
within twice the stellar half mass radius. The $f_g$ ranges were 0.01--0.03, 0.03--0.1,
0.1--0.3 and 0.3--1.  

Figure 1 shows the gas mass fraction distributions of Illustris galaxies
compared to IllustrisTNG galaxies in 4 different ranges of stellar mass. As in KBN16, both the gas
and the stellar mass are evaluated within twice the stellar half-mass radius.
There has been no attempt to split the total gas into atomic, molecular and ionized
components as in Diemer et al (2018) Stevens et al (2018)  -- the comparison here is between the two simulations. An
attempt to assess the match with observations using a sample of 
galaxies with both atomic and molecular gas measurements  is presented in Figure 1 of KBN16, where
it was found that the simulations yield too many very gas-rich galaxies at
Milky Way masses  (3 $\times 10^{10} -10^{11} M_{\odot}$).

\begin{figure}
\includegraphics[width=121mm]{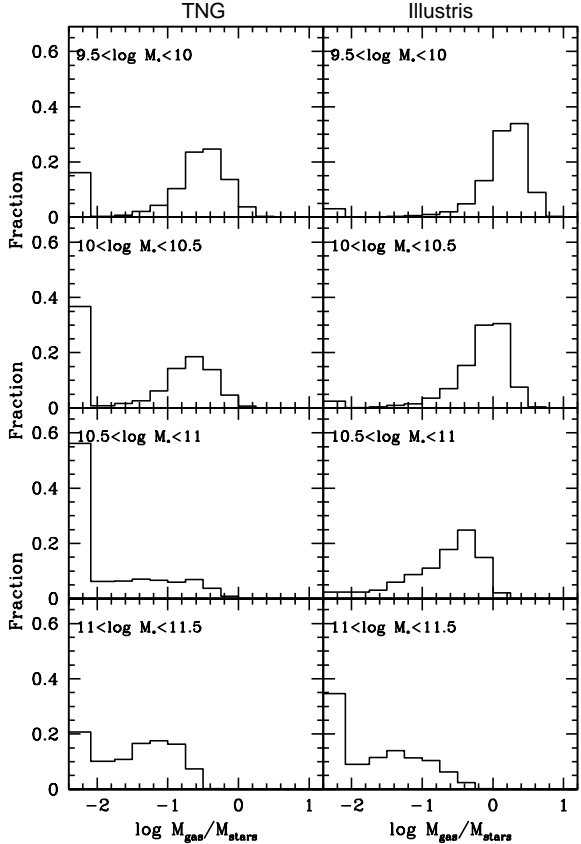}
\caption{ Black histograms show total gas mass fractions evaluated within
twice the stellar half mass radius for galaxies in the IllustrisTNG (left) 
and Illustris (right) simulations.
Results are shown in four different stellar mass bins. 
\label{models}}
\end{figure}

As can be seen, the gas mass fraction distributions in the two simulations differ most
strongly in this mass range;
the fraction of galaxies with gas mass fractions less than 0.01
reaches values in excess of 0.5 in IllustrisTNG compared to only a few percent in
Illustris, and there are very few galaxies in IllustrisTNG with gas mass
fractions close to 1.  This is because almost all galaxies in this stellar mass
range have black holes with masses of around $10^8 M_{\odot}$. 
In IllustrisTNG, AGN kinetic feedback kicks in very strongly 
at this black hole mass and AGN feedback prescriptions are thus maximized for Milky Way mass systems
(see Weinberger et al 2018). We thus regard the 3 $\times 10^{10} -10^{11} M_{\odot}$
stellar mass range as ideal for looking for observational tests of the
differences in the feedback prescriptions.
At the same time, we caution that a  black hole mass of 
$10^8 M_{\odot}$ is too large for disk-dominated galaxies
like the Milky Way
by a factor of 3 to 10 (Genzel, Eisenhauer \& Gillesen 2010). This would imply
that the effects of black hole feedback are simply 
too dominant in TNG Milky Way-type galaxies. 

The strong differences at the Milky Way  mass scale also cause
problems for making robust systematic  comparisons between Illustris and
IllustrisTNG as a function of galaxy gas mass fraction. 
In order to sample a wide range in gas mass fraction in both simulations,
we adopt slightly
different stellar mass and halo mass cuts for the two.  For
Illustris, we select galaxies with stellar masses in the range $10.6<\log
M_*<10.8  M_{\odot}$ and halo masses in the range $12.0<\log M_{halo}<12.3$ ,  very similar
to the cut in KBN16.  For IllustrisTNG, we lower the stellar and halo mass
limits slightly, and select galaxies with stellar masses in the range $10.4<\log
M_*<10.7  M_{\odot}$ and halo masses in the range $11.7<\log M_{halo}<12.3 M_{\odot}$.  We only carry
out the comparison for 3 bins in $f_g$: 0.03–-0.1, 0.1–-0.3 and 0.3–-1, for which
there are more than 20 subhalos in each bin in both simulations.  We note that
galaxies in the highest gas fraction bin will be slightly less massive in
IllustrisTNG than in Illustris, but we believe that the adopted mass ranges are
similar enough that the comparison is still meaningful.
We also note that the stellar-mass-to-halo mass relation is modified in TNG
compared to Illustris, (see Figure 11 of Pillipich et al (2018)), with TNG
lower at fixed halo mass at the mass scale of Milky Way-type galaxies, so
the offset adopted here is within the systematic differences in the 
calibration of the two simulations.

\subsubsection {Matched halo analysis} In order to understand differences in the
dynamics (inflow/outflow) and thermodynamic state of the CGM in the two
simulations, we apply a matching algorithm to identify pairs of analogue
subhaloes across the two simulations. 
Counterparts are identified 
using the Lagrangian region matching algorithm of Lovell et al. (2014). The
initial conditions for a Lagrangian patch of each halo is determined using the
dark matter particles that the subhalo has at the snapshot at which it achieves
its maximum mass. The Lagrangian patch is then compared to the patches of haloes
in the companion simulation by means of their density distributions and
gravitational potentials to obtain a quality-of-match statistic. The companion
simulation subhalo with the highest value of this statistic is then considered
the ``match''.  We have selected a set of 169 matched subhalos using method (2)
with masses in the range $6 \times 10^{11} - 2 \times 10^{12} M_{\odot}$ that
contain galaxies with stellar masses in the range $5 \times 10^{10} - 10^{11}
M_{\odot}$ with gas mass fractions larger than 0.01.

The full sample is shown in the left panel of Figure 2. We plot the gas fraction
of the galaxy in IllustrisTNG as a function of the gas fraction of its Illustris
analogue. In the right panel, we define a set of subsamples for further study:
1) a set galaxies with $f_{gas}> 0.3$ in both Illustris and IllustrisTNG
(hereafter the gas-rich sample), 2) a set of galaxies with $0.1 < f_{gas} <0.3$
in both simulations (hereafter, the moderate gas sample), 3) a set of glaxies
with $f_{gas} < 0.1$ in the IllustrisTNG simulation and $f_{gas}>0.1$ in
Illustris (hereafter, gas-poor TNG galaxies). We note that we could
have defined  a set of galaxies with
$f_{gas}<0.1$ in both simulations, but further examination revealed
that all such cases in
Illustris are actually satellite subhalos.
In these systems, environmental processes such as ram-pressure
stripping  affect the CGM in addition to processes such as cooling, supernovae
and AGN feedback.  Because of this extra degree of complexity, we do not consider
galaxies in this regime of parameter space.

\begin{figure}
\includegraphics[width=91mm]{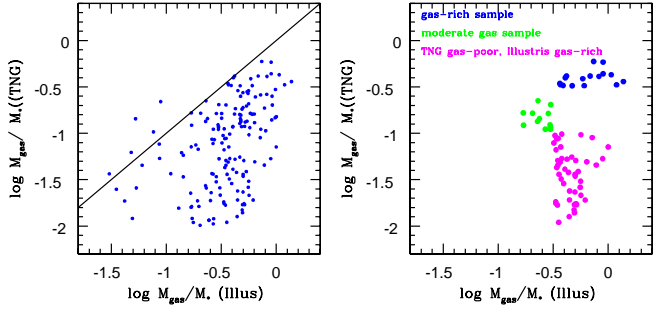}
\caption{Left: The full sample of matched subhalos. The gas fraction
of the galaxy in IllustrisTNG is plotted as a function of the gas fraction of its Illustris
analogue. A one-to-one line is drawn to guide the eye. 
Right: The subsamples that are studied in this paper (see text).}    
\label{models}
\end{figure}

\section {CGM structure trends as a function of galaxy gas fraction}
In this section, we present a comparison of the radial profiles of the gas, as
well as probes of gas morphology, asymmetry and the coherent rotation of the CGM
for galaxies as a function of gas mass fraction in Illustris and IllustrisTNG.
Computational methodologies, results for the Illustris simulation and some
comparison with available observational data were discussed in detail in KBN16,
and the reader is referred to sections 3.2-3.5 and 4 of that paper for details.
Here, we focus on the comparison between the two models.

Figure 3 shows the mean radial distribution of the logarithm of the total gas density, neutral
gas column density and gas mass weighted temperature as a function of radius from the
center of the subhalo. Results are shown for IllustrisTNG in the left panels and
for Illustris in the right panels.  The red, green and blue curves show results
for the samples with $f_g$ in the ranges 0.03--0.1, 0.1--0.3 and 0.3--1,
respectively.  The gas masses, neutral hydrogen fractions and gas temperatures
for each cell are read directly from the simulation outputs.  As can be seen,
the radial distributions of all three quantities depend strongly on galaxy gas
mass fraction out to radii of 70-100 kpc in the Illustris simulation. Gas-rich
galaxies are surrounded by a higher density, cooler CGM on average than gas-poor
galaxies. The same is not true, however, in IllustrisTNG. The average densities
and temperature of the CGM on scales larger than 10 kpc do not depend on the gas
fraction of the galaxy.

\begin{figure}
\includegraphics[width=90mm]{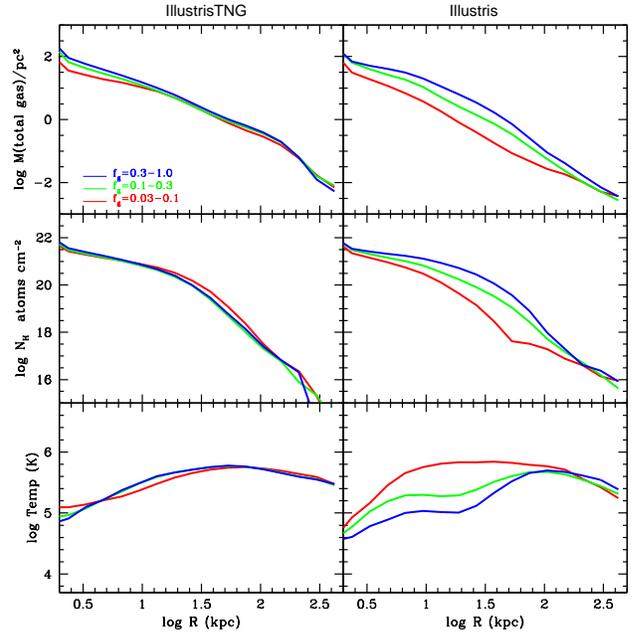}
\caption{The mean radial distribution of the logarithm of the total gas density, neutral
gas column density and gas mass weighted temperature as a function of radius from the
center of the subhalo. Results are shown for IllustrisTNG in the left panels and
for Illustris in the right panels.  The red, green and blue curves show results
for the samples with $f_g$ in the ranges 0.03--0.1, 0.1--0.3 and 0.3--1, respectively.}
\label{models}
\end{figure}

Figure 4 explores the extent to which the CGM gas is aligned positionally with
the disk of the galaxy. Different colour lines show radial profiles
of neutral hydrogen column density  evaluated at
different orientation angles with respect to the major axis of the disk.
Results are shown for IllustrisTNG in the left column  and for Illustris in the right
column. Results for galaxies in different gas fraction ranges are plotted in
different rows.  As discussed in KBN16, there is no clear alignment of the CGM
neutral gas in the plane of the disk in the Illustris simulation for the more
gas-rich systems.  In IllustrisTNG,  we do see a clear trend for the average
neutral gas density to be larger when measured in the plane of the disk. The
systematic boost in $\log N_H$ reaches a factor of 10 and is present out to
distances of 100 kpc.

\begin{figure}
\includegraphics[width=90mm]{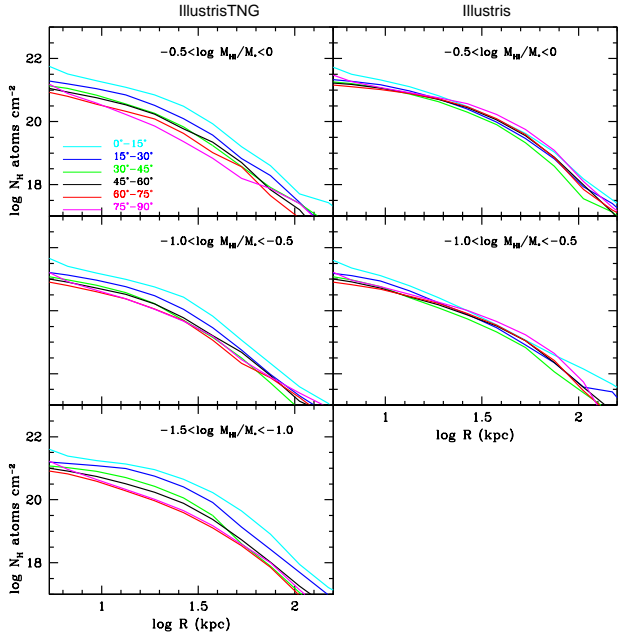}
\caption{Different colour lines show radial profiles 
of neutral hydrogen column density  evaluated at
different orientation angles with respect to the major axis of the disk:
0$^{\circ}$ --  15$^{\circ}$ (cyan), 15$^{\circ}$ –- 30$^{\circ}$ (blue),
30$^{\circ}$ –- 45$^{\circ}$(green), 45$^{\circ}$ -- 60$^{\circ}$(black),
60$^{\circ}$ – --75$^{\circ}$(red), 75$^{\circ}$ –- 90$^{\circ}$(magenta).  Results are
shown for IllustrisTNG in the left column  and for Illustris in the right
column. Results for galaxies in different gas fraction ranges are plotted in
different rows.}
\label{models}
\end{figure}

In KBN16, we defined a set of asymmetry indices that we designed to probe the
average asymmetry of the CGM, both above and below the galactic plane, and on
either side of the disk. To measure these indices, the galaxy was first rotated
into its edge-on configuration. Radii along the major axis $R_x$(90), $R_x$(95)
and $R_x$(99) enclosing 90, 95 and 99 percent of the in-plane projected stellar
mass were measured. The asymmetry indices A$_c$ (up-down 90), A$_c$ (up-down
95), A$_c$ (up-down 99), A$_c$ (right-left 95) and A$_c$ (right-left 99), were
defined, where ``up-down'' means above and below the galactic place, and
``right-left'' means on either side of the disc minor axis. To compute the
indices, we integrate up the total gas columns from a scale height of 2 kpc
above/below the plane out to a distance of 200 kpc, and compute (for example)
A$_c$ (up-down 90) as $\log ([M_{tot}(\rm{up})−-M_{tot}(\rm{down})]/
[M_{tot}(\rm{up})+M_{tot}(\rm{down})])$, i.e. the index is
expressed in terms of the logarithm of the fractional difference in total gas
mass above and below the central disc. The other four indices are defined in
similar fashion.

Figure 5 compares the distributions of some of these indices in the two
simulations.  Results for IllustrisTNG are shown in the top row and in the
bottom row for Illustris. Red histograms are for gas-poor galaxies with
$f_g=0.03-0.1$, while blue histograms are for gas-rich galaxies with
$f_g=0.3-1$. In Illustris, gas-poor galaxies have more asymmetric neutral gas
distributions than gas-rich galaxies.  The differences in asymmetry are much
smaller for the total gas. As discussed in KBN16, the asymmetric neutral gas
distributions are the effect of the bubble feedback, which acts to heat the gas
and leads to irregular holes in the neutral gas distribution. In IllustrisTNG,
there is no difference in CGM asymmetry between gas-rich and gas-poor galaxies.

\begin{figure*}
\includegraphics[width=169mm]{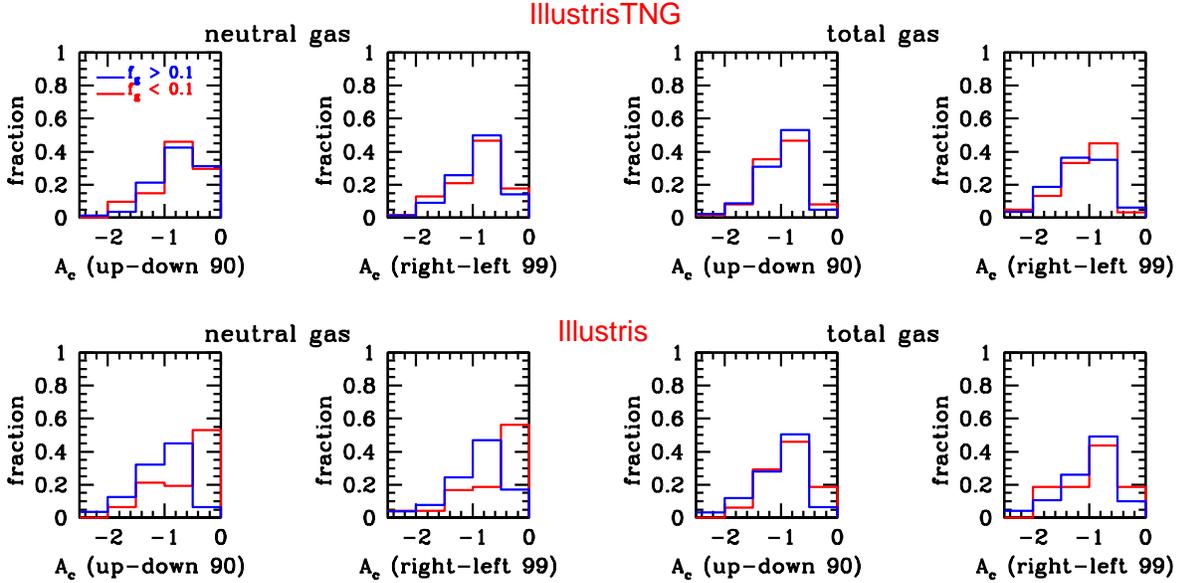}
\caption{ Comparison of  2 different asymmetry indices for the neutral gas
distribution and the total gas distribution. 
Results are shown for gas-rich
($f_g>0.1$; blue histograms) and gas-poor ($f_g<0.1$; red histograms) galaxies.
The two indices represent 
the asymmetry in the central gas distribution above
and below the plane and
the asymmetry in the far outer disk to the right and to the left of the disk minor axis.
Results for IllustrisTNG are shown in the top panels and for Illustris in the bottom
panels.
\label{models}}
\end{figure*}

Finally, we explore the rotational coherence scale of the CGM.  Once again, we
rotate the galaxy into an edge-on configuration and find the positions along the
major axis, $x_{min}$ and $x_{max}$ where the measured velocity reaches its
maximum and minimum values.  The full width half-maximum (FWHM) of the vertical
coherence length of the rotation is estimated by centering at $x_{min}$ and
$x_{max}$ and finding the distance y over which the velocity is greater than
half its minimum/maximum value. The distributions of FWHM vertical extent of
coherent rotation are plotted in Figure 6 for the IllustrisTNG and Illustris
gas-rich (blue) and gas-poor (red) subsamples. As can be seen, the FWHM values
span the same range of values for gas-rich galaxies in the  two simulations. The main difference is
that there is a systematic shift towards smaller FWHM values for gas-poor
galaxies in Illustris, but not in IllustrisTNG. Again, in KBN16 this trend was
attributed to more vigorous bubble feedback in gas-poor galaxies, 
which acts to destroy coherent structures of neutral gas.

\begin{figure}
\includegraphics[width=91mm]{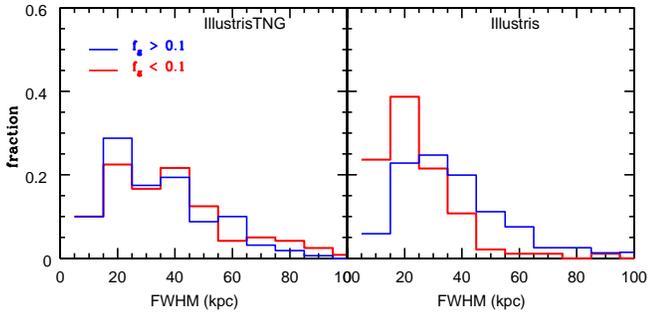}
\caption{Histograms of the fraction of galaxies as a function of FWHM vertical extent
for gas-poor galaxies with $f_g < 0.1$ (red) and for gas-rich galaxies with
$f_g > 0.1$ (blue). Results are shown for IllustrisTNG in the left panel and
for Illustris in the right panel. 
\label{models}}
\end{figure}

In summary, we find that large scale CGM properties in IllustrisTNG such as mean
column density, asymmetry and rotational coherence scale, depend much more
weakly on the galaxy gas fraction than they do in Illustris. The neutral gas
distribution in IllustrisTNG is also preferentially aligned in the plane of the
central disk, whereas it is much more isotropic in Illustris.

\section {Tracer particle analysis of inflowing/outflowing and
heating/cooling gas}

The equations of hydrodynamics are  solved in astrophysical applications using
one  of two general approaches: particle-based Lagrangian-like schemes such as
Smoothed Particle Hydrodynamics (SPH), or mesh-based Eulerian-like schemes.
Lagrangian-like SPH schemes have an  advantage that it is possible to follow
fluid resolution elements in time and track the evolution of their properties,
at the expense of introducing errors on the smoothing scale (Vogelsberger et al 2012, Section 5.3).
In Eulerian-like schemes the discretized quantity is the volume itself, and
information on the past evolution of the fluid is lost.  Genel et al (2014)
developed a tracer particle scheme in the moving-mesh code AREPO (Springel
2010). The scheme is based on a Monte Carlo sampling of fluid motions. Tracer
particles are attached to fluid cells and whenever two cells exchange mass, they
also exchange the appropriate fraction of their tracer particles. In this way,
the tracer particles can be used to follow gas flows as a function of time in
the simulation, with limitations due to Monte Carlo statistical noise (Nelson et al 2013).

In this section, we compare gas temperature evolution and gas flows in Illustris
and IllustrisTNG using the sample of matched subhaloes described in the previous
section. The matching technique allows us to divide the full sample into a set
of sub-samples according to galaxy gas fraction and still carry out meaningful
comparisons using sets of around a dozen subhalos. Our aim is to understand how
the changes in the sub-grid models regulating supernovae and AGN feedback impact
the physical state of the CGM.

\subsection {Cooling/heating and inflow/outflow balance} 
We begin with an examination of the balance between cooling and heating, as well
as inflow and outflow in the two simulations. We work with all the gas tracer
particles located within the subhalos in our samples at z=0 and extract the 
positions, velocities and temperature of all these particles at a series of
earlier output times.   We
first recenter our coordinate system to the center of mass of the stars in the
subhalo at z=0. We select a region around the center of mass within which the
stellar density is larger than 2\% of the  central stellar density and then
rotate the galaxy into its edge-on configuration defined by the stellar
particles. The x-axis of the new coordinate system is defined to lie along the
major axis of the edge-on system, while the z-axis is perpendicular to the
galactic plane. We work with the gas tracers at 3 different snapshot times
corresponding to z=0.1 ($t_{look-back}$= 1.3 Gyr), 0.5($t_{look-back}$= 5.1 Gyr),
1 ($t_{look-back}$= 7.8 Gyr).  For each of these snapshots, we again recenter the
tracer particles to the center of mass of the stars and rotate them into the
edge-on configuration of the inner stellar distribution at that snapshot.  In
this way, the bulk motion of the galaxy between the two snapshots is removed,
and the motion of the gas between the two snapshots is defined with respect to
the centre of mass of the disk.  We define the change in radius $\Delta R$ as
$\sqrt{x_1^2+y_1^2+z_1^2}- \sqrt{x_0^2+y_0^2+z_0^2}$, where $(x_0,y_0,z_0)$ are
the coordinates of the tracer particles at z=0 and $(x_1,y_1,z_1)$  are the
coordinates at the earlier snapshot. If the radius has decreased between the two
snapshots, we label the tracer as inflowing, and if the radius has decreased, we
label it as outflowing. Tracer particles also carry the thermodynamic properties
of the gas and we calculate $\Delta T$ as the difference in temperature between
the two snapshots. \footnote {We note that cooling of gas is not tracked accurately below a 
temperature of $\sim 10^4$ K in the simulations.  We have simply used the snapshot
values at face value, disregarding the implementation of the Springel \& Hernquist (2003)
subgrid model for the interstellar medium. This means that $\Delta$T values very close to
the disk should be disregarded.}  If the temperature has increased, the tracer is labelled as
heating, and if it has decreased, the tracer is labelled as cooling.
Note that the longer the adopted look-back time, the more likely it
is that the  
tracers  have complex  behavior
within the two time windows. For gas in a fountain
flow, the tracer particles could have been outflowing, then inflowing, then outflowing again,
between two of our tracer time points. The quantities plotted should thus be regarded
as a measure of the {\em net} flow averaged over many tracers within the given time window.

In the upper panels of Figure 7, we plot the logarithm of the ratio of cooling
versus heating tracer particles as a function of the distance of the particle
from the center of the subhalo at z=0. In the lower panels, we plot the
logarithm of the ratio of inflowing versus outflowing tracer particles as a
function of radius. Note that the  mean numbers
of tracer particles are equivalent to ``mass'', because every tracer represents
a constant baryon mass.
Results for IllustrisTNG are shown in the left column and in
the right column for Illustris. Red, green and blue curves are for the gas-poor,
moderate gas and gas rich subsamples (see section 2.2). No gas-poor results are shown
for Illustris, because these are all 
satellite galaxies in this simulation. Solid curves show
results evaluated between z=0.1 and z=0 (i.e over a look-back time of 1.3 Gyr),
while dashed curved show results evaluated over a look-back time of 8 Gyr, from
z=1 to z=0.

\begin{figure}
\includegraphics[width=91mm]{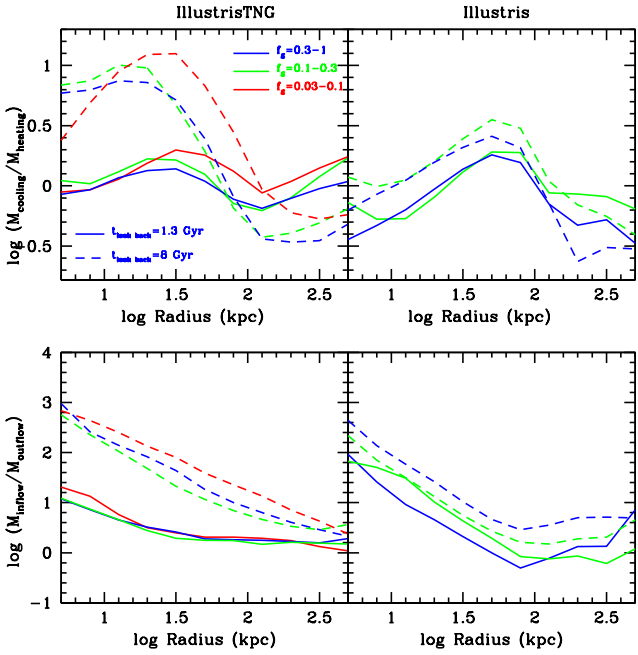}
\caption{ Upper panels: logarithm of the ratio of cooling
versus heating tracer particles is plotted as a function of the distance of the particle
from the center of the subhalo at z=0. Lower panels: logarithm of 
the ratio of inflowing versus outflowing tracer particles is plotted as a
function of radius. Results for IllustrisTNG are shown in the left column and in
the right column for Illustris. Red, green and blue curves are for the gas-poor,
moderate gas and gas rich subsamples. Solid curves show
results evaluated between z=0.1 and z=0 (i.e over a look-back time of 1.3 Gyr),
while dashed curved show results evaluated over a look-back time of 8 Gyr, from
z=1 to z=0.} 
\label{models}
\end{figure}

In IllustrisTNG, the CGM is in approximate heating/cooling equilibrium over the
past 1.3 Gyr with $M_{cooling} \simeq M_{heating}$ all the way from the inner
disk out to a radius of 100 kpc.  The inflowing mass is larger than the
outflowing mass out to a radius of around 30 kpc, but the CGM at larger radii is
in inflow/outflow equilibrium. When evaluated over a timescale of 8 Gyr, the
picture changes quite dramatically. The cooled mass dominates over the heated
mass by a factor of 10 in the inner 30-50 kpc of the halo, and approximate
heating/cooling equilibrium is only reached at radii greater than 100 kpc.
Likewise, the mass of gas that has flowed in over the last 8 Gyr is larger than
the mass that has flowed out at all radii out to a few hundred kpc.

In Illustris, the dependence of the results on the adopted look-back time is much
weaker.  This means that the heating/cooling and inflow/outflow patterns in the
CGM are largely {\em time-independent} out to z=1. The inner CGM is inflow
dominated, but in rough equilibrium  between cooling and heating. The ratio of
cooling mass over heating mass increases with radius, reaching a maximum at R=70
kpc, before dropping sharply in the outer (R$>100$ kpc) region of the halo.
Likewise, the region of the halo where the inflowing mass dominates over the
outflowing mass ends at R$\sim$ 100 kpc, which likely 
indicates the region of the halo where
the bubble feedback is heating the gas and pushing most of it outwards.

In summary, the results in Figure 7 indicate a  progessive decrease in the amount of 
cooling and inflowing gas around IllustrisTNG Milky Way mass galaxies from z=1 to z=0.
This occurs across the entire subhalo, but most strongly in the
inner regions.  In
contrast, the feedback in Illustris is {\em time-independent}, always acting to
quench the inflow of cooling gas in the outer regions of the halo.  Note that
these results appear to be largely independent of the present-day gas fraction
of the galaxy in both simulations.

The next two figures  examine how the {\em total gas mass in the halo} tagged as
heating or cooling (Figure 8) and as inflowing or outflowing (Figure 9) is
partitioned as a function of radius. The format of these two figures and the
meanings of the different line colours and styles in the same as in Figure 7
(see captions for more details).In IllustrisTNG, the mass distributions tagged
according to whether the tracer particles have been cooling or heating  appear
to be bimodal, with one part of the distribution located  between 10 and 100 kpc
and another one located  between 100-400 kpc. The heated gas contributes mainly
to the lower radius peak, while cooling gas is mainly located at larger
distances. In Illustris, the radial separation  between gas that is heated and
cooled is not as pronounced.  Compared to IllustrisTNG, there is a clear
suppression of gas at the very largest radii, once again indicative of the
effect of the bubble feedback model in pushing gas at large radii out of the
subhalo. Similar conclusions are reached when the tracers are tagged according
to whether they are inflowing or outflowing (Figure 9).  Illustris exhibits
somewhat stronger separation in radius between the inflowing and outflowing
components than between the cooling and heating components, while the opposite
is true for IllustrisTNG.

Another way to think about the results presented in these two figures is
in terms of the differing size-scales of the fountain flows (or of the 'baryon
cycle' established by the feedback mechanisms) in the two simulations. In
IllustrisTNG, there are two main flow patterns: one that is established on scales
of a few tens of kpc that is maintained by a combination of supernova/AGN
feedback and gravitational  infall onto the central disk, and another one
that is established on scales of a few hundred kpc that is maintained by
cosmological infall and shock-heating of gas. In Illustris, the outer
flow is suppressed in fractional mass contribution compared to the inner flow.   

\begin{figure}
\includegraphics[width=91mm]{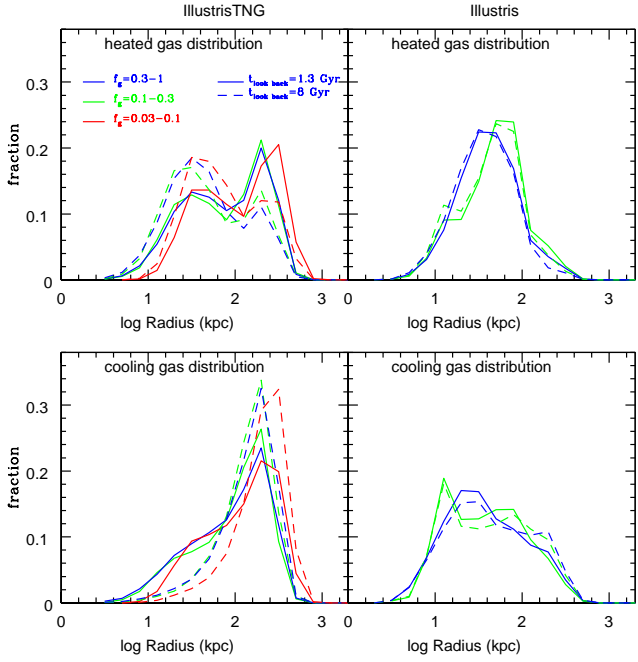}
\caption{ The distribution of the the total gas mass in the halo tagged as
heating or cooling as a function of radius. The format of the figure and the
line styles and colours have the same meaning as in Figure 7.}
\label{models}
\end{figure}

\begin{figure}
\includegraphics[width=91mm]{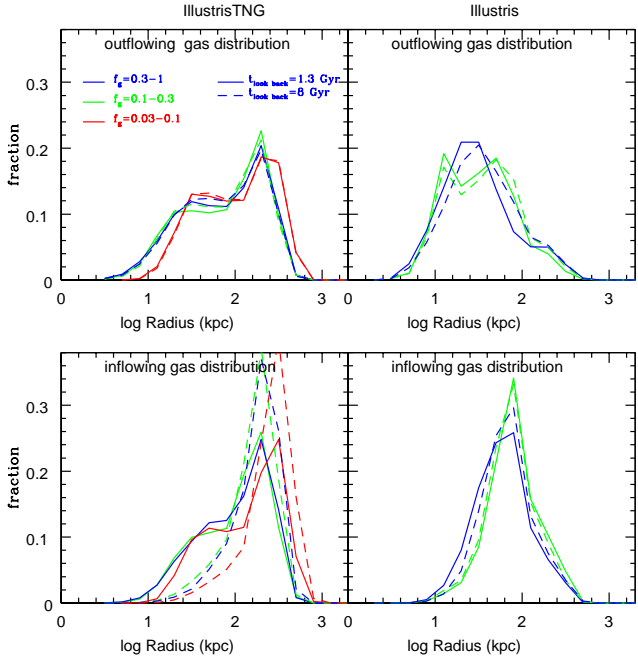}
\caption{ The distribution of the the total gas mass in the halo tagged as
inflowing or outflowing as a function of radius. The format of the figure and the
line styles and colours have the same meaning as in Figure 7.}
\label{models}
\end{figure}

In summary, we have examined the balance between heating and cooling and inflow and
outflow as  a function of radius in the CGM. We have uncovered significant
differences between the two simulations, indicative of the differing
implementations of feedback in the two cases. 

\subsection {Gas accretion onto the disk}
We now turn to a study of the nature of the gas that accretes
onto the disk.

After rotating the galaxy into its edge-on configuration and recentering to the
center of mass of the inner stellar distribution, we define the gas disk as the
radial region $r< R_{max}$  in the disk plane  within which the gas surface
density does not drop below $1/e^2$ of its central value. We then calculate
height above and below the disk plane, $z_{min}$ and $z_{max}$   where the gas
density stays above $1/e^2$ of the average value within the gas disk. The region
$r< R_{max}$ and $z_{min} < z < z_{max}$ is defined  as the  ``accreted zone''.
Tracer particles (either stars or gas) found within this zone at z=0, which are
located outside it at an earlier snapshot in the form of gas, are defined as
accreted particles. In Figure 10, we plot the
temperature
distribution of these tracers at earlier times, as well as their radial
distribution at the earlier snapshots.
The histograms have been normalized so that
the sum over all bins is equal to the fraction of the z=0 disk
mass that has been accreted.    Once again,
results are shown for the two simulations in different columns.  Solid curves
show results for a look-back time of 1.3 Gyr and dashed curves for a look-back
time of 8 Gyr.  Red, green and blue curves indicate results for the gas-poor,
moderate gas and gas-rich sub-samples, respectively.

\begin{figure}
\includegraphics[width=91mm]{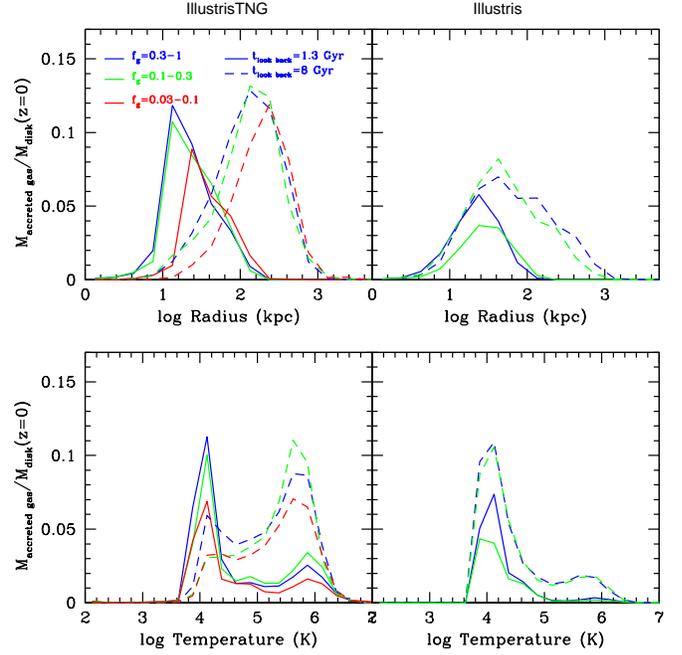}
\caption{ The radial and temperature distribution of gas particles at redshifts
$z=0.1$ (solid) and $z=1$ (dashed) accreted onto the disk at $z=0$. 
The format of the figure and the
line styles and colours have the same meaning as in Figure 7.}
\label{models}
\end{figure}

In both Illustris and IllustrisTNG, gas-rich galaxies have undergone more
accretion than gas-poor galaxies. In Illustris, this trend is only seen for
recent look-back times, but in IllustrisTNG, this persists for look-back times out
to z=1. We also see that in Illustris, the majority of the accretion is in the
form of relatively cool gas, when evaluated over long look-back times. In
contrast, in IllustrisTNG, recent accretion is from cool gas located at radii
between 10 and 100 kpc from the disk but when evaluated over look-back times of 8
Gyr, we see that much of the accreted material was hot with temperatures $\sim
10^6$ K at z=1 and that at this redshift, this gas was located between 100 kpc
and 1 Mpc from the disk. We note that the trend in CGM accreted fraction with
galaxy gas fraction is quite small; this implies that the subsequent conversion rate of
gas into stars in the disk is likely playing a significant role in setting the
observed present-day gas fraction of the galaxy in both simulations.

To summarize, 
Figures 8, 9 and 10, show that accretion in Illustris is dominated by a
small-scale (10-50 kpc) fountain flow of mainly cold material.
This is consistent with the conclusions of Nelson et al (2015b), who showed that
accretion in Illustris is dominated by recycled material out to redshifts $\sim 1$.
In TNG, however, the longer timescale picture is
accretion of hotter gas from larger distances. This implies a
different balance between  hot gas cooling from the halo  and cool, recycled fountain
gas as the
accretion source for the galaxy. 

The strong evolution of cooling/accretion with redshift  
in TNG compared to Illustris can likely be attributed to
different redshift evolution of  wind properties. 
The right panels of Figures 6 and 7 of  Pillepich et al (2018) show that
the evolution of
the wind mass loading factor and velocity at injection with
redshift actually have
different behaviour.
In dark matter halos with masses $10^{12} M_{\odot}$ in TNG, the wind velocity
at injection remains constant at a value of $\sim 1000$ km/s from z=4 to z=0 
in TNG, while the wind mass loading factors decrease by a factor of 2 from
4 to 2 over this redshift range. This implies that SN feedback evolves
only weakly with redshift.
The  likely explanation for the strong evolution in CGM
properties with redshift in this simulation is strongly evolving AGN feedback,
as the galaxy transitions into the kinetic mode at the threshold mass.   
In Illustris,  the wind velocity decreases from 700 km/s at
z=4 to 400 km/s at z=0, while the 
wind mass loading factors increase from 2 to 4.5 over the same redshift range.
These opposing trends, combined with the fact that AGN feedback affects gas
mainly at large distances in the halo, may offer an explanation for why CGM
properties in this simulation are
so invariant with redshift.  

\subsection {How does the CGM differ between gas-rich and gas-poor galaxies?}

It is reasonable to suppose that way to pinpoint why some galaxies
are gas rich and some galaxies are gas-poor is by separating the CGM gas into
different thermodynamic and kinematic components as a function of radius. 

In Figures 11 and 12,
we show how tracer particles in the  CGM are partitioned within the space of
$\Delta R$ versus $\Delta T$, where $\Delta R$ is the change in radius of the
particles, with negative values indicating inflow and positive values indicating
outflow, and $\Delta T$ is the change in temperature, with negative values
indicating cooling and positive values indicating heating. Results for the inner
CGM (radii less than 30 kpc) are shown in Figure 11 and for the outer CGM (radii
greater than 30 kpc) are shown in Figure 12. We have adopted a tracer time interval of
5 Gyr for the inner CGM and 8 Gyr for the outer CGM.  In both figures, IllustrisTNG
results for three gas fraction subsamples are presented in the left columns,
while the right columns show results for the two Illustris subsamples. The contour
spacing in each panel is 0.4 dex in the logarithm of the fraction of the total
number of particles; the lowest contour level (plotted in black) corresponds to
$\log$ F=-4.9 and the highest contour level (plotted in white) corresponds to
$\log$ F=-1.2.

\begin{figure*}
\includegraphics[width=169mm]{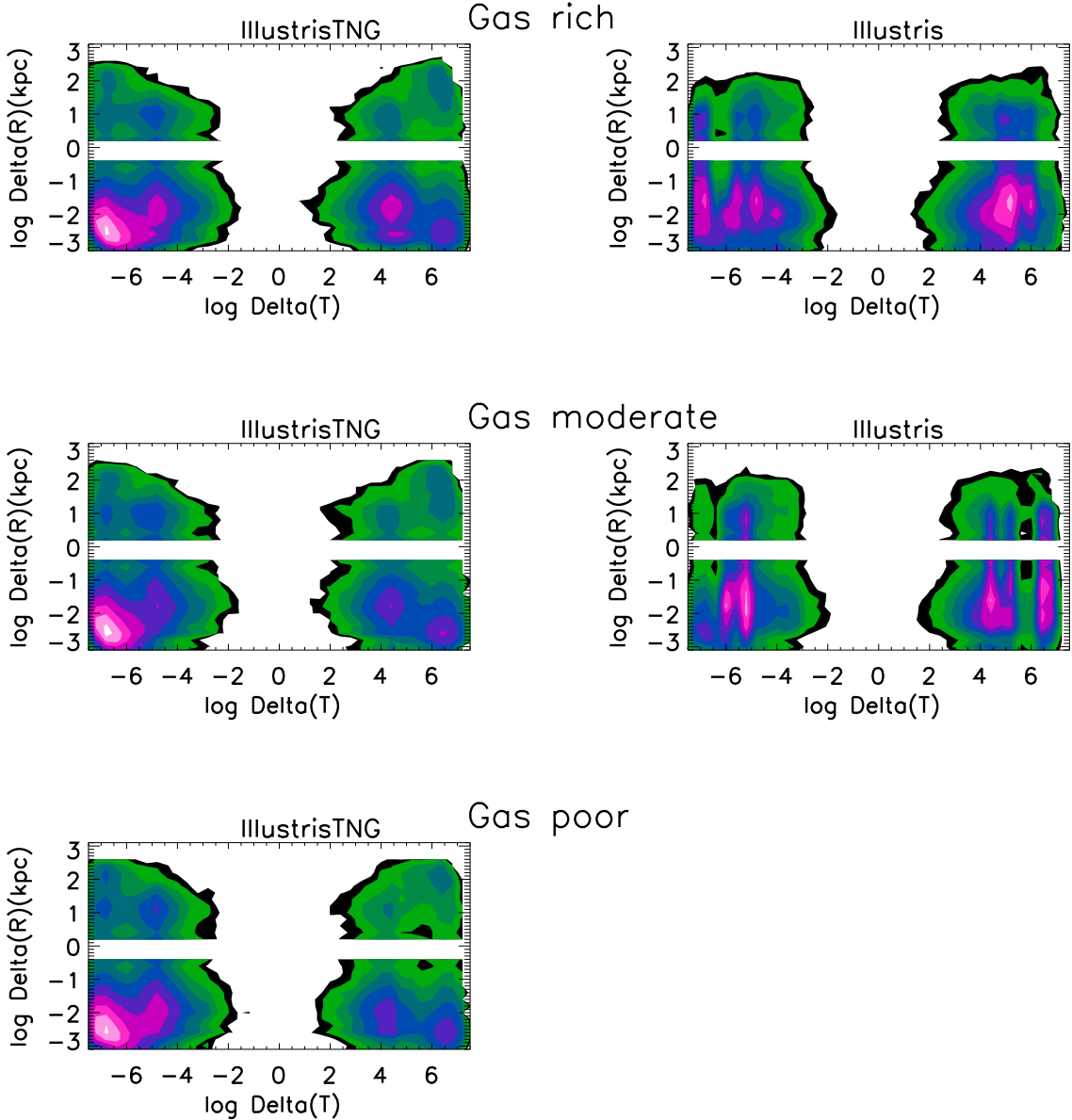}
\caption{ The distribution of 
 tracer particles in the  CGM within 30 kpc from the center
of the subhalo in the 2-dimensional plane of  
$\Delta R$ versus $\Delta T$, where $\Delta R$ is the change in radius of the
particles, with negatve values indicating inflow and positive values indicating
outflow, and $\Delta T$ is the change in temperature between $z=0.5$
and $z=0$, with negative values
indicating cooling and positive values indicating heating. The contour 
spacing in each panel is 0.4 dex in the logarithm of the fraction of the total
number of particles within the radius; the lowest contour level (plotted in black) corresponds to
$\log$ F=-4.9 and the highest contour level (plotted in white) corresponds to
$\log$ F=-1.2.  Results are shown for three  different gas fraction subsamples
for IllustrisTNG and two different ones for Illustris.}
\label{models}
\end{figure*}

\begin{figure*}
\includegraphics[width=169mm]{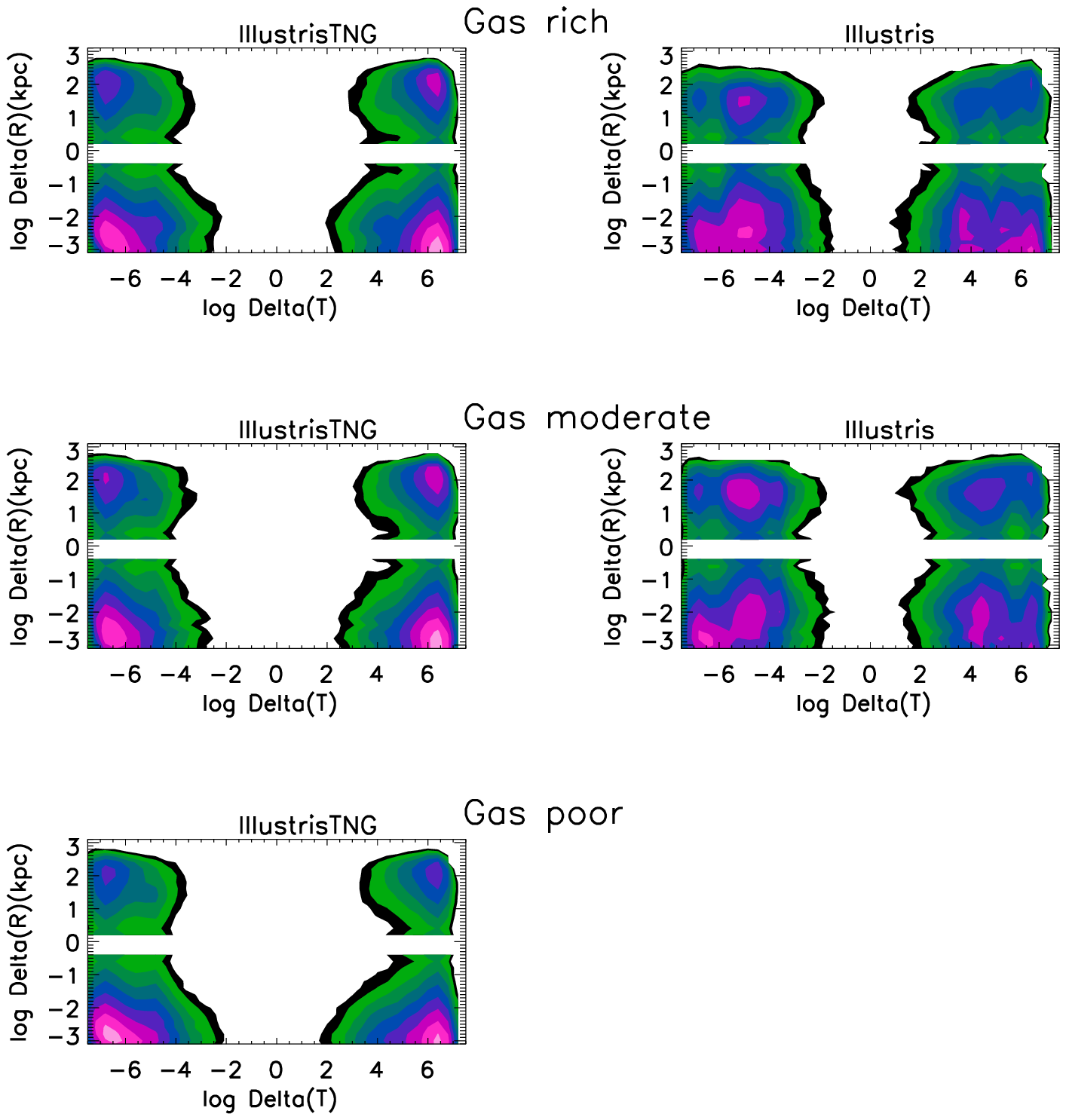}
\caption{ As in the previous figure, except for
 tracer particles in the  CGM further way than 30 kpc from the center
of the subhalo. $\Delta R$ and  $\Delta T$ are evaluated between $z=1$ and $z=0$.  }   
\label{models}
\end{figure*}

In each panel in the two figures, there are four distinct quadrants
corresponding to, a) cooling gas that is inflowing (lower left), b) cooling gas
that is outflowing (upper left), c) heated gas that is inflowing (lower right),
d) heated gas that is outflowing (upper right). The sum over all 4 quadrants is
renormalized to 1, so that comparisons between  different panels  can be made in
a meaningful way. Symmetry between positive and negative values on the $\Delta$T and
$\Delta$R axes 
would indicate that the gas is in thermal and inflow/outflow  equilibrium, respectively.  
In agreement with results shown in Figure 7, 
we see that on small scales, gas is strongly
inflowing and cooling in IllustrisTNG. In Illustris, gas is inflowing but is
in rough thermal equilibrium.   On large scales, the inflowing gas is in thermal
equilibrium in both simulations, as expected since most gas 
is virialized within the halo and cooling times are long.  
The outflowing gas component, on the other hand,  is experiencing net heating 
in IllustrisTNG , but net cooling in Illustris, which presumably reflects the
different subgrid models for feedback in the two simulations.

In IllustrisTNG, the most striking change as a function of
the gas mass fraction of the galaxy  is that there is a clear increase in the
fraction of gas present in the heated, inflowing component on small scales for
more gas-rich systems.  In Illustris, the most striking change in the CGM
surrounding galaxies of different gas mass fraction occurs in the outer regions.
Gas poor systems have more mass (both material that is cooling and being heated)
in an outflowing component.

To summarize, we find that the reservoir that varies most with the gas fraction of a
galaxy is distributed  over very different scales in the two simulations. In IllustrisTNG, the reservoir
is in the form of heated, galactic fountain gas on scales of
a few tens of kpc that later re-accretes onto
the central system.  In Illustris, the feedback energy is dumped at larger
distances from the central system, and the gas fraction is regulated by how much
of the large-scale ($\sim$ 100 kpc)  gas reservoir is able to cool and re-accrete
over long timescales .  

\subsection {Example of a gas-rich galaxy in both simulations} 
To illustrate the different gas components
in a pictorial fashion, we  select matched subhalos in the gas-rich
sub-samples of Illustris and IllustrisTNG, and plot the tracer particles.
Figures 13 (IllustrisTNG) and 14 (Illustris) show the tracers  in  edge-on and
face-on configurations  for one typical matched subhalo in the gas-rich
subsample.  The upper panels show the inner region of the disk, limited to a
radius of 20 kpc at z=0, while the lower panels show a larger scale view of
where the same particles were  located at z=0.5. The tracer particles have
been colour-coded according to whether they have experienced heating/cooling or
inflow/outflow between z=0.5 and z=0: blue indicating particles that are cooling
and inflowing, green indicating particles that have been heated but have
flowed into the central disk, yellow indicating material that has cooled 
but is outflowing, and red indicating material that is heated and outflowing.

\begin{figure*}
\includegraphics[width=169mm]{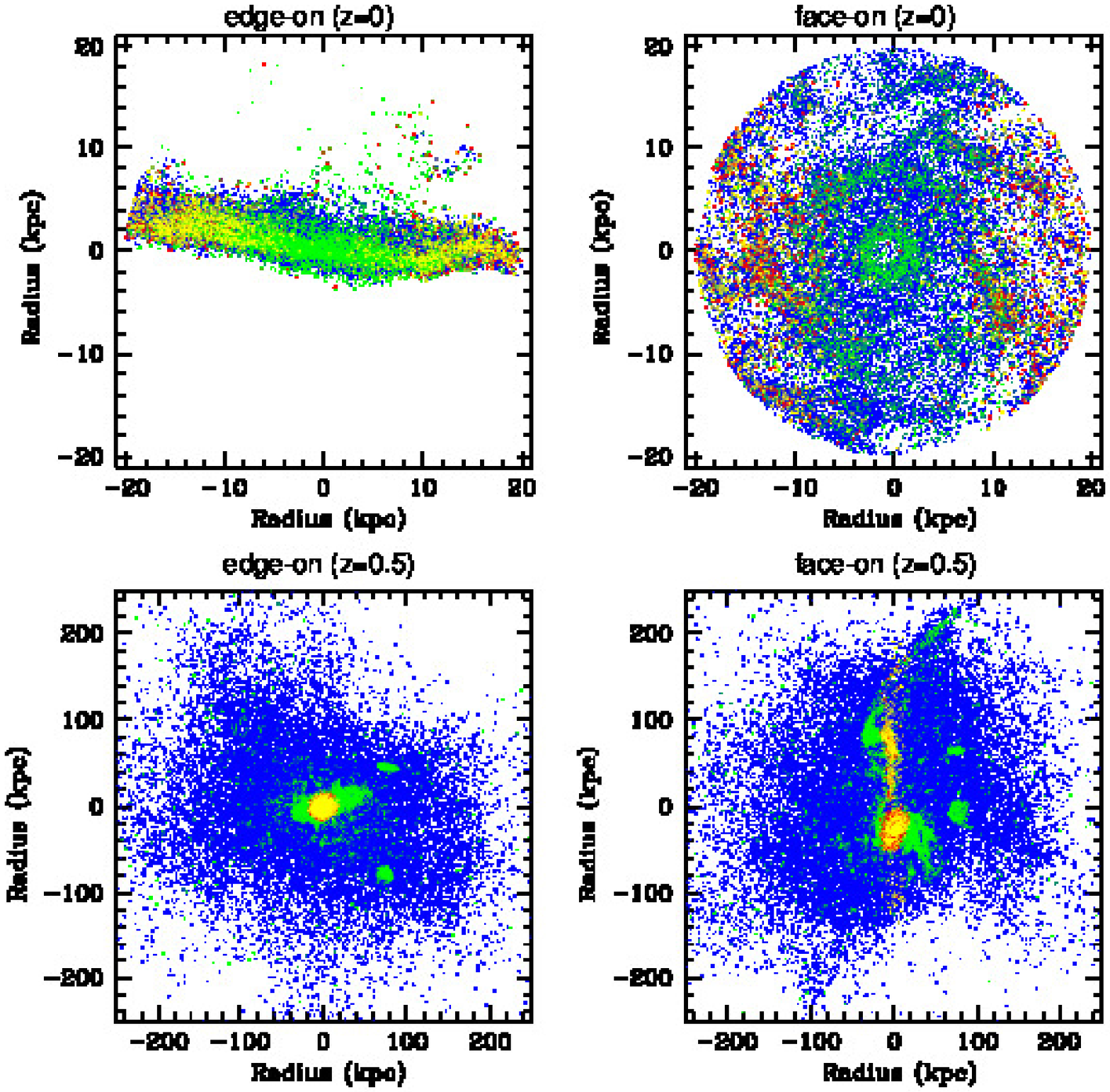}
\caption{ Tracer particles in one gas-rich galaxy in IllustrisTNG are plotted in  edge-on and
face-on configurations.  
The upper panels show the inner region of the disk, limited to a
radius of 20 kpc at z=0, while the lower panels show a larger scale view of
where the same particles were located at z=0.5 The tracer particles have
been colour-coded according to whether they have experienced heating/cooling or
inflow/outflow between z=0.5 and z=0: blue indicating particles that are cooling
and inflowing, green indicating particles that have been heated but have
flowed into the central disk, yellow indicating material that has cooled 
but is outflowing, and red indicating material that is heated and outflowing.} 
\label{models}
\end{figure*}

\begin{figure*}
\includegraphics[width=169mm]{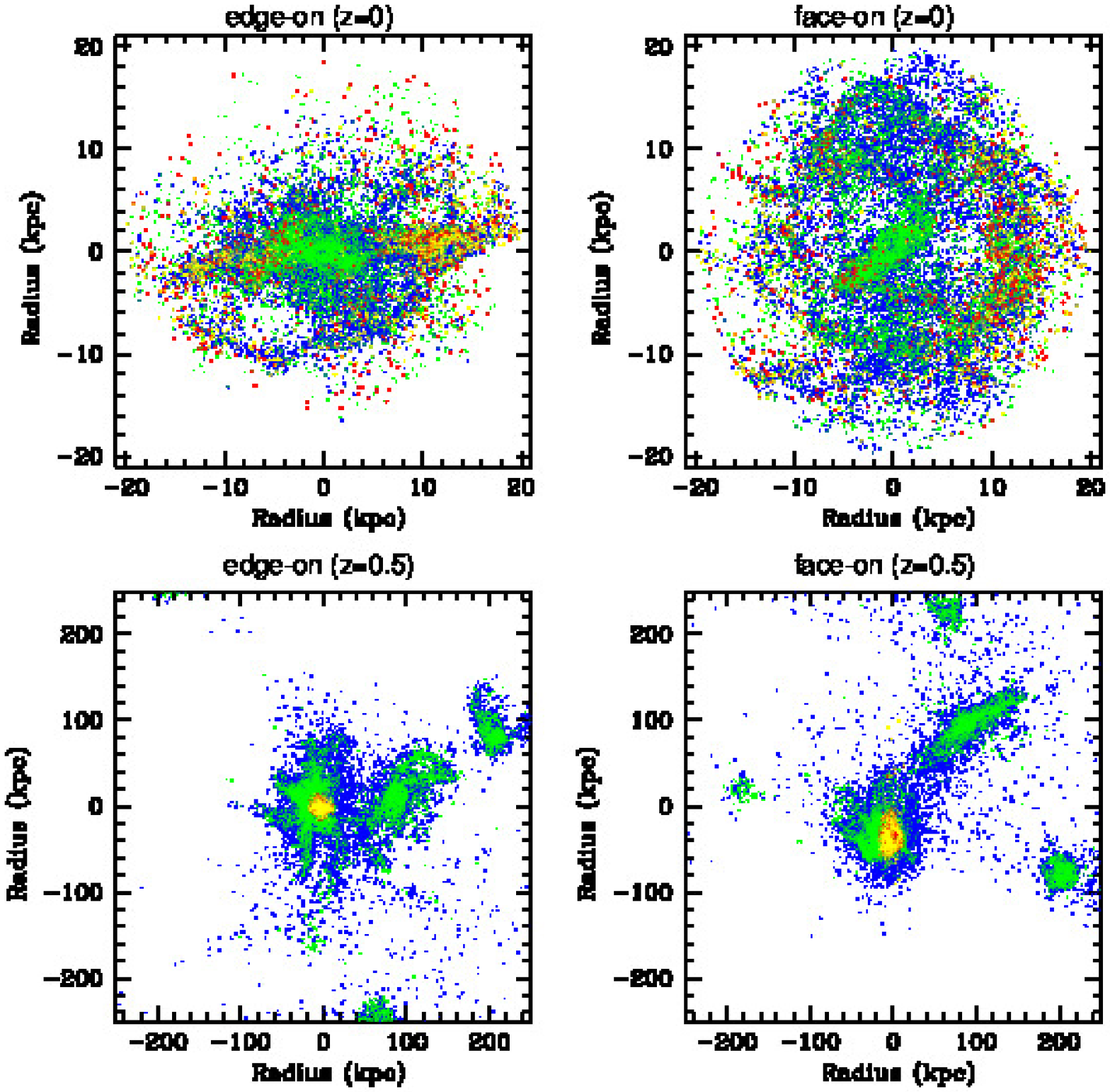}
\caption{ As in Figure 13, for the tracer particles in the matched subhalo in Illustris.} 
\label{models}
\end{figure*}

There are a number of very striking differences between the simulations, which
we will now discuss. The reader should recall that matched subhalos originate
from the same patch of particles in the initial conditions of the two
simulations and therefore the dark matter merger histories will be close to
identical, since gravitational forces are dominated by the dark matter
component and not the baryons.  The differences between Figures 13 and 14
should largely reflect the changes in the prescriptions for supernova and AGN feedback in the
two simulations, modulo any 'timing offset' caused by the fact that 
the two systems might be in slightly different evolutionary stages in the
two runs as a result of the differences in the treatment of
the baryonic physics and the small difference in
cosmological parameters in the two simulations. 
Comparison of the top two panels in the two figures shows that
the gas is much more extended perpendicular to the disk in Illustris, in
agreement with our finding in Section 3, that Illustris gas-rich disks have
large vertical coherence lengths. The distribution of the gas away from the disk
midplane is not uniform, but exhibits large bubble-like structures, which are
likely tidal debris remnants rather than AGN feedback signatures, because most
of the gas in the vertically extended part of the disk is cold and inflowing.
In both simulations, the heated gas is always located in denser-than-average
structures in the disk, indicating that the heated gas traces regions where
stars are forming more rapidly. The heated, outflowing gas is located at larger
radius in the disk than the heated, inflowing gas, as expected. In
Illustris, heated, outflowing gas also contributes to the vertically extended
gas component at the largest scale heights above the disk plane .

Moving to the lower panels of Figures 13 and 14, we see that in IllustrisTNG, a
significant fraction of the disk has cooled and condensed from a smooth, rather
uniform reservoir of material extending out to radii of around 150 kpc.  Most of
the  heated, inflowing gas is located in a flattened structure of around 80 kpc
in diameter. A smaller fraction is located in satellite systems that are
destined to merge with the main disk.  The outflowing gas is the most centrally
concentrated. In contrast, the gas distribution at z=0.5 in Illustris is
contained in 3-4 much more massive structures and there is very little smooth
reservoir. The gas disk has thus grown through a series of merging events, which
explains why the present-day disk is considerably more irregular than in
IllustrisTNG and also exhbits a strong bar.

In summary, in the most gas-rich galaxies in IllustrisTNG, feedback processes
have acted at high redshifts to drive a significant fraction of the gas into a
smooth, quasi-spherical halo-like structure, which then condenses to form a thin
disk at low redshifts. In IllustrisTNG, much of the gas is accreted recently in
in the form of clouds of recycled material, which merge in a 
configuration that creates a thickened,
rotating structure.

\section {Summary and Future Perspectives}

The goals of this analysis were two-fold: a) to understand the extent to which
the conclusions drawn in KBN16 about trends in structure and kinematics of the
CGM as a function of galaxy gas mass fraction are dependent on the adopted
prescription for supernovae and AGN feedback in the simulations, b) to
understand how structural and kinematic trends reflect underlying physical
processes at work in the CGM, namely cooling and inflow of gas onto the galaxy
under gravity as the surrounding dark matter halo assembles over time, as well
as heating and outflow of gas as a result of energy ejection by stellar and
black hole sources.

The answer to the first investigation is that all the trends identified in KBN16
are no longer present in IllustrisTNG. For our sample of Milky Way mass halos at z=0,
gas column density  and temperature
profiles at large radii ($\sim 100$ kpc) no longer correlate with the gas mass
fraction of the central galaxy in IllustrisTNG. The neutral gas at large radii
is preferentially aligned in the plane of the disk in IllustrisTNG, whereas it
is much more isotropic in Illustris. There is also no longer any relation
between asymmetries in the neutral gas distribution and galaxy gas mass
fraction. Finally, the vertical coherence scale of the rotationally supported
gas in the CGM is also no longer linked to the gas mass fraction of the galaxy.

Our tracer particle analysis allows us to construct a fairly detailed physical
picture to explain why the differences between the simulations are so striking.
In IllustrisTNG, there is a very strong redshift dependence in the properties of
the CGM. The CGM  at z=1 is very strongly cooling and inflow dominated, and then
evolves to a state where heating/cooling and inflow/outflow rates roughly
balance each other at z=0.  In contrast, there is little redshift evolution of
CGM properties out to z=1 in Illustris.

When we look in detail at the morphology of the material that has cooled to form
the galactic disk in Illustris, we find that it is located in  a smoothly
distributed, isotropic halo of gas that has cooled from temperatures of $\sim
10^5-10^6$ K at z=1 to $10^4$ K at the present day. In contrast, the material
that has formed the disk in Illustris is already in the form of lumps of cool
($\sim 10^4-10^5$ K) gas at z=1.  These results can be understood by considering
the very different way in which feedback operates in the two simulations. In
IllustrisTNG, kinetic feedback operates  efficiently near the centres of galaxies
at high redshifts to drive gas out of these systems (i.e. the winds have high
mass-loading factors).  This reservoir of smooth material cools and collapses to
form a thin, rotating disk by the present day. In Illustris, the feedback energy
is dumped at large radii by heating the gas in large "bubbles", which pushes a
significant fraction of the hot gas out of the halo. The cooler lumps merge,
forming a more vertically thickened structure.

Finally, when we look at how the gas is partitioned into inflowing/outflowing
and heating/cooling components as a function of radius around galaxies of
different gas mass fractions, we are able to understand in more detail how the
CGM is regulating the gas supply to the disk in both simulations. In
IllustrisTNG, the regulation to the disk occurs on smaller scales than in
Illustris.  Gas-rich galaxies have a larger fraction of inflowing heated gas on
scales of a few tens of kpc that maintains the gas supply to the galaxy. In
contrast, the galaxy gas fraction in Illustris depends on the degree to which
bubble feedback has managed to empty the halo on larger scales.  In short,
IllustrisTNG galaxies are fuelled by ``galactic fountains'', while the gas
supply for Illustris galaxies is regulated on much larger scales by processes
that heat and remove the gas from their halos.

Perhaps the most striking and unexpected result from this work is the degree to
which the low redshift CGM is dependent on the detailed implementation of
supervova and AGN feedback processes in the simulated galaxies, most of which
reach their peak efficiency at redshifts greater than 1. Much effort is now
being invested in characterizing outflows from high redshift star-forming and
active galaxies (e.g. Davies et al 2018); our results suggest that checks on
observed parametrizations of velocities and mass-loading factors of outflowing
gas from star-forming galaxies (Nelson et al 2019, in preparation) should be complemented by
systematic characterization of the structure and kinematics of gas around low
redshift galaxies to ensure that a consistent picture emerges from the
simulations. As discussed in KBN16, this can be done in a number of ways,
including quasar absorption line statistics (Tumlinson et al 2013; Werk et al
2014; Borthakur et al 2015), radio interferometric studies of HI around galaxies
(e.g. Wang et al 2014),  deep observations of extra-planar, diffuse ionized gas
using integral field unit spectra (e.g. Jones et al 2017) and X-ray and
Sunyaev-Zeldovich studies of hot gas halos around galaxies (e.g. Anderson et al
2015). 

The fact that no systematic trends in CGM properties are found as a function of
galaxy gas mass fraction in IllustrisTNG in contradiction with the results of
Borthakur et al (2015) would appear to indicate that further tuning of the feedback
recipes will be required to match observations in detail.  
Kauffmann, Borthakur \& Nelson (2016) showed that even though the observed 
disk gas fraction distributions agree well with Illustris,
the covering fraction of neutral gas is a factor of two smaller
than the data at large radii in the halos. This is attributed to the radio
mode feedback heating gas to high temperatures at large radii from the
central source. Nelson et al (2018) show that the gas covering fraction
problems discussed in this paper 
are alleviated in IllustrisTNG. The remaining issue is whether
the strong alignment of neutral gas with the HI disk is inconsistent with
data. The main problem is that the  Borthakur absorber
sample currently consists of only 45 galaxies that have stellar masses
uniformly distributed between $10^{10}$
and $10^{11} M_{\odot}$. By construction, the feedback physics in the simulations
changes strongly across this mass range. In all simulations, this is
necessary in order to match the transition from a predominantly blue
population of galaxies at $10^{10}$ M$_{\odot}$  to a predominantly red population of galaxies at 
$10^{11} M_{\odot}$. 
In future, analysis of larger galaxy samples and  continued
dialogue between observations and simulations will hopefully yield a reasonably
complete answer to the puzzle of how galaxies acquire their gas.


\end{document}